\documentclass[%
 reprint,
 amsmath,amssymb,
 aps,
]{revtex4-2}

\usepackage{graphicx}
\usepackage{dcolumn}
\usepackage{bm}
\begin{document}

\preprint{APS/123-QED}

\title{Capillary-driven thinning of DNA solutions}

\author{Vincenzo Calabrese$^{1,*}$}

\author{Silvia Nardone$^{1, 2}$}
\author{Amy Q. Shen$^1$}
\author{Simon J. Haward$^1$}

\affiliation{$^1$Micro/Bio/Nanofluidics Unit, Okinawa Institute of Science and Technology Graduate University\\1919-1 Tancha, Onna-son, Okinawa 904-0495, Japan}%
\affiliation{$^2$DICMaPI, Università degli Studi di Napoli Federico II, P. le Tecchio 80, 80125 Napoli, Italy}%

\email{vincenzo.calabrese@oist.jp}
\email{simon.haward@oist.jp}






\date{\today}

\begin{abstract}
Capillary thinning of polymeric fluids is central to biological and industrial processes, yet the mechanisms governing thinning dynamics remain unresolved, especially for semi-flexible polymers. Using ideal solutions of semi-flexible DNA, we validate a predictive model for exponential capillary thinning that accounts for each polymer in the molecular weight distribution. For semi-flexible polymers, self-selection of the exponential time constant occurs by a fundamentally different mechanism than for highly flexible systems, and is not simply governed by the longest polymer relaxation time.
\end{abstract}

\maketitle

Sufficiently slender fluid filaments self-thin under the action of surface tension. This phenomenon known as capillary-driven thinning is ubiquitous in both biological and industrial processes involving fluid breakup \cite{McKinley2005, keshavarz2016ligament,ruhs2021complex,zinelis2024transition}. For instance, in the printing industry, capillary-driven thinning drives the formation of ink droplets, while during sneezing or coughing, it dictates the size of mucus droplets, affecting the transmission of viral diseases \cite{dhand2020coughs}.
During capillary-driven thinning of fluid filaments, the radial squeezing of the filament by surface tension generates a uniaxial extensional flow. For Newtonian fluids (water, honey), the physics of capillary-driven thinning is well understood, as thinning is controlled by a balance of surface tension against viscous and inertial forces \cite{Papageorgiou1995,McKinley2005,McKinley2000}. For more complex fluids such as polymer solutions, additional elastic forces may come into play, and the filament thinning dynamics becomes intimately linked to the orientational dynamics of the polymer in the extensional flow field. These coupled dynamics lead to the attractive possibility of extracting rheological properties (viz. characteristic polymer ``relaxation times’’) from straightforward time-resolved measurements of the width of the thinning polymeric filaments \cite{McKinley2005, Campo2010, dinic2020flexibility, Gaillard2023, gaillard2024does, zinelis2024fluid, aisling2024importance}. However, significant questions remain over how the coupling between thinning and polymeric dynamics depends on factors such as polymer stiffness or flexibility, architecture, concentration in solution, molecular weight, and polydispersity of the molecular weight.


\begin{figure}[b!]
    \centering
    \includegraphics[width=8cm]{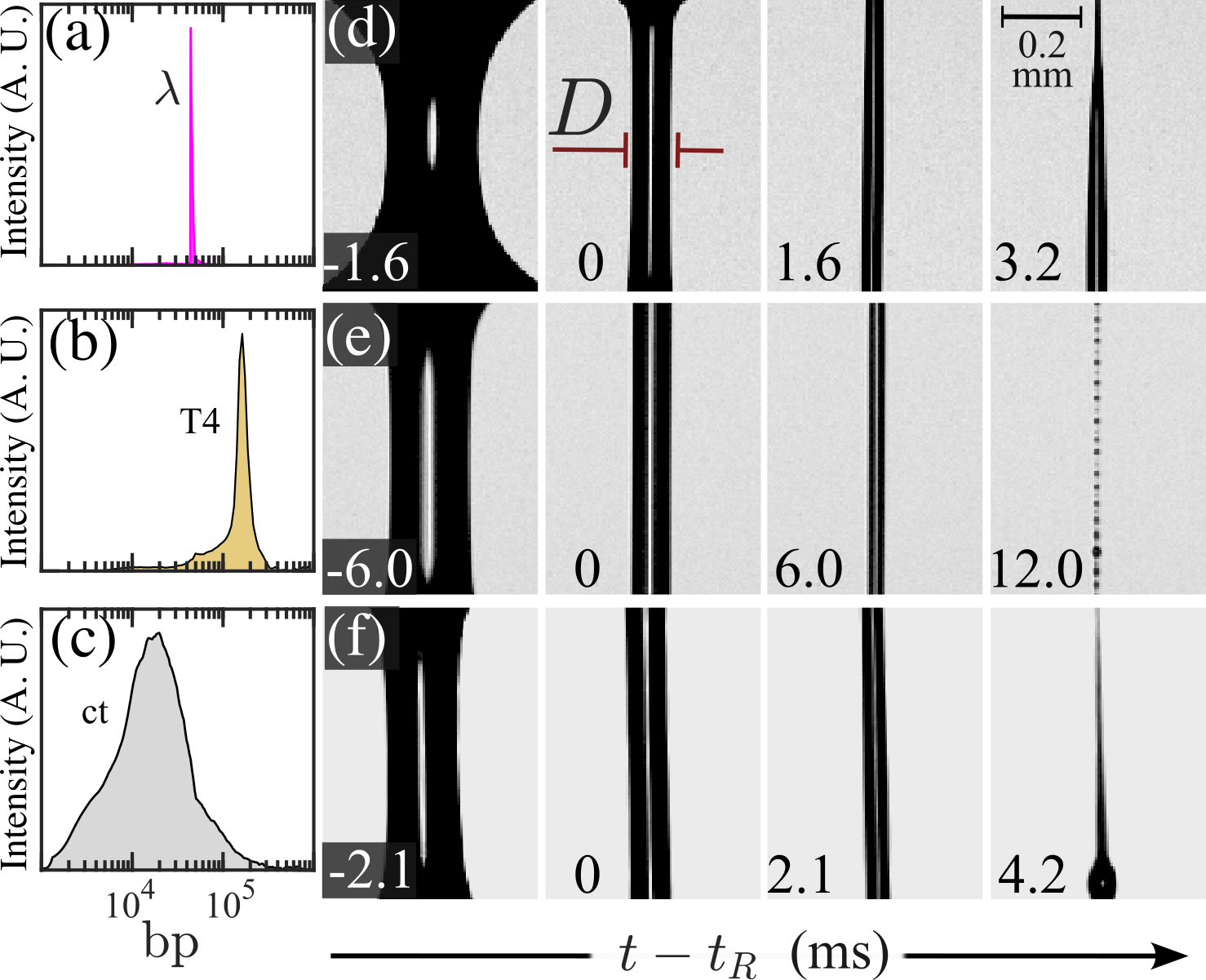}
    \caption{Base pair (bp) distribution for (a) $\lambda$, (b) T4 and (c) ct DNA. (d-f) Snapshots of the capillary-driven thinning of DNA solutions. (d) $\lambda$ at $c/c^*=1.8$, (e) T4 at $c/c^*=1.7$, and (f) ct-DNA at $c/c^*=2.2$. The reference time $t_R$ is taken as the time $t$ when the filament diameter $D=0.1$~mm.}
    \label{fig:Images}
\end{figure}
Strong extensional flow developed during the capillary-driven thinning of polymeric fluids leads to stretching and alignment of polymer chains and the appearance of an elastocapillary (EC) regime \cite{Bazilevsky1990,Entov1997, clasen2006dilute,sattler2007molecular,juarez2011extensional}. The EC regime is characterized by an exponential decrease of the diameter at the neck of the filament, $D$, with time, $t$, according to the time constant $\tau_{EC}$ as,
\begin{equation}
D \sim \mathrm{exp}(-t/ \tau_{EC}).
\label{eqn.1}
\end{equation}
The most controversial aspect is how the fluid self-selects the elastocapillary time constant, $\tau_{EC}$, based on polymer conformation and concentration. The prevailing interpretation is based on the Oldroyd-B constitutive model, which assumes polymer chains to be infinitely extensible.
This model asserts that $\tau_{EC}$ is determined by the longest polymer relaxation time, $\tau\equiv \tau_{EC}/3$. However, it is increasingly evident that this relationship does not hold universally \cite{calabrese2024polymers, Gaillard2023, zinelis2024fluid,gaillard2024does}. Experimental and numerical studies have shown that the Oldroyd-B model becomes progressively less accurate in describing polymer dynamics in the EC regime as polymer stiffness increases \cite{zinelis2024fluid, calabrese2024polymers}. Thus, for semi-flexible polymers (e.g., polyelectrolytes and most biopolymers) cases where $\tau_{EC}/3 \ll \tau$ have been reported \cite{calabrese2024polymers}. 
Despite intense experimental efforts to study semi-flexible polymers in capillary-driven flows, these studies have provided limited understanding due to the use of polydisperse polymers with unknown size distributions and uncertain average molecular weight ($Mw$) \cite{dinic2020flexibility,Jimenez2018,Jimenez2020,calabrese2024polymers,sattler2007molecular}. 

\begin{figure*}[t!]
    \centering
    \includegraphics[width=17cm]{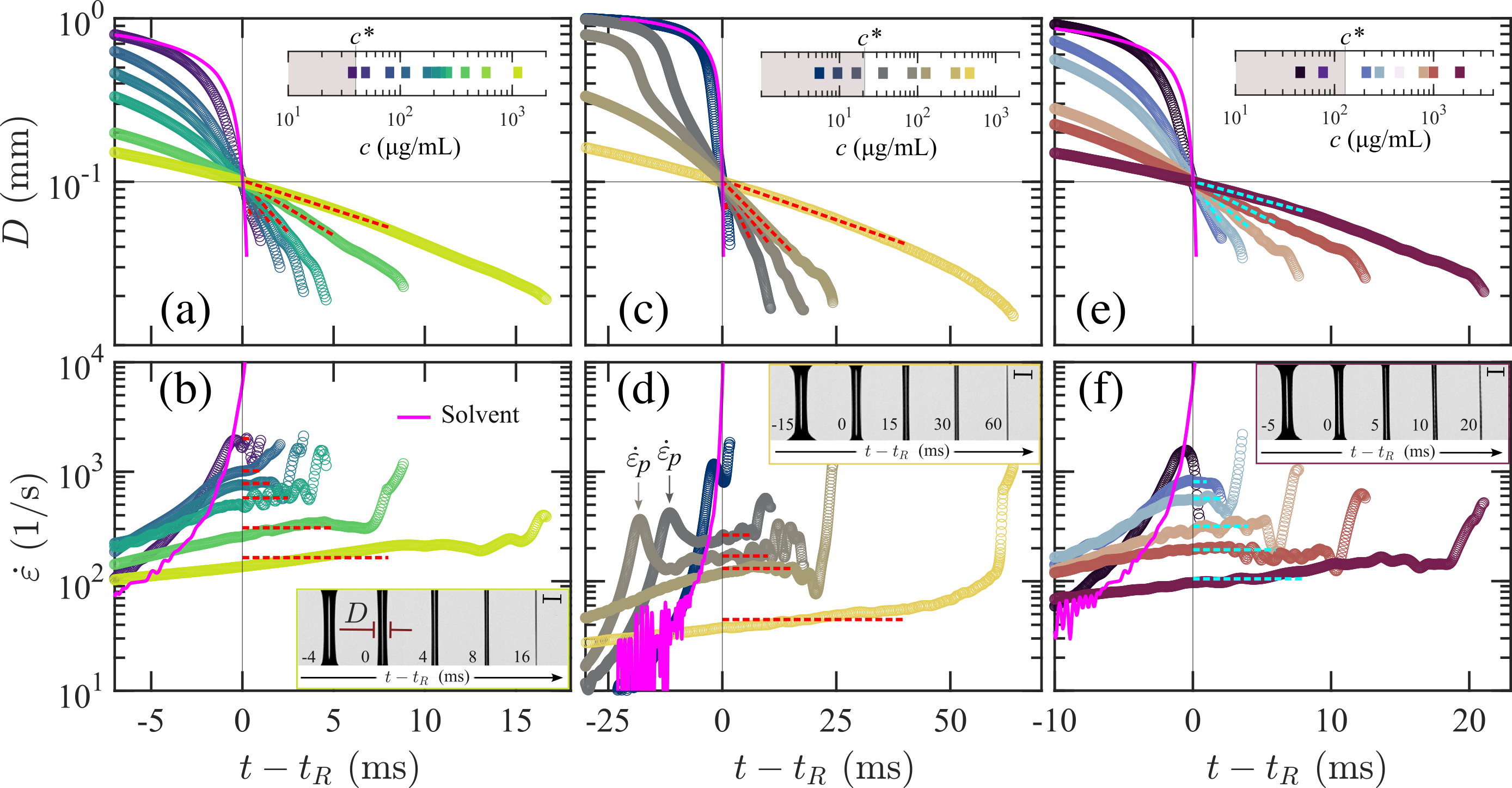}
    \caption{Filament diameter $D$ and extension rate $\dot\varepsilon$ as a function of time $t-t_R$ (top and bottom raw, respectively) for the $\lambda$ (a,~b), T4 (c,~d) and ct-DNA (e,~f). The dashed lines in (a, c, e) indicate the fit to eqn. \ref{eqn.1} and in (b, d, f) the corresponding $\dot\varepsilon_{EC}
    $. For clarity, only representative fluids are shown. The insets in (b), (d), and (f) show snapshots of the respective DNA solutions at the highest measured concentrations, scale bar is 0.2~mm.  In panel (d), the peak extension rate $\dot\varepsilon_p$ is highlighted.}
    \label{fig:2}
\end{figure*}

In this Letter, we tackle this problem by employing monodisperse DNAs as a model semi-flexible polymer to establish a fundamental understanding of how $Mw$ and polymer concentration influence $\tau_{EC}$. To extrapolate our findings to more complex systems, we also investigate well-characterized bidisperse and polydisperse DNA solutions. We propose a simple model that captures the experimental observations and highlights that, in solutions of semi-flexible polymers, the time constant $\tau_{EC}$ accounts for the time scale contribution from each DNA size present in the population.


We use monodisperse $\lambda$-DNA (Thermo Fisher) and T4-DNA (Nippon Gene) with $48.5\times 10^3$ and $166.0 \times 10^3$ base pairs (bp), corresponding to $Mw \simeq 32$ and 108 MDa, respectively. As polydisperse DNA we use calf thymus DNA, ct-DNA (MP Biomedicals) with a peak size of $20 \times 10 ^3$ bp ($Mw \simeq 13$ MDa). The bp-distribution of the tested DNAs was measured using a femto pulse electrophoresis system (Agilent Technologies) (Fig.~\ref{fig:Images}(a-c)). The samples are prepared by dilution of the DNA stock solutions with a Newtonian aqueous buffer (Tris-EDTA 10 mM Tris-HCl, 1 mM EDTA, pH 8.0), following previous guidelines \cite{harnett2024effects, banik2021monodisperse} (see Supplemental Material for details \cite{ESIref}). For each sample, the DNA concentration $c$ and sample homogeneity is determined using a UV-Vis spectrophotometer (Nanodrop, Thermo Fisher) \cite{banik2021monodisperse}. For $\lambda$ and T4-DNA, the overlap concentration $c^*=40$ and 21~\textmu g/mL, respectively, is estimated based on simple cubic packing fraction of DNA coils \cite{rubinstein2003polymer, brunet2016dependence, reisner2012dna, ESIref}. For polydisperse ct-DNA $c^*=128$ ~\textmu g/mL is estimated experimentally from rheometric measurements \cite{ESIref}. 

The self-driven capillary thinning of the fluid filament is measured at $22^\circ$C using the low-inertia slow retraction method (SRM) implemented on a commercial capillary breakup extensional rheometer (CaBER) device (Haake, ThermoScientific) \cite{Campo2010}. The fluid is loaded in the 1~mm gap between two coaxial circular plates (with diameter $6$~mm). The lower plate is moved downwards using a micro-adjustable stage causing the fluid to undergo capillary-driven thinning. The filament thinning is recorded using a high-speed camera (Phantom Miro 310 at 17000 frame/s) and the minimum filament diameter $D$ is retrieved using a MATLAB routine following the guidelines given in ref. \cite{ng2021highlighting}. 
Fig~\ref{fig:Images}~(d-f) shows snapshots of the capillary-driven self-thinning for $\lambda$, T4 and ct-DNA fluids at a comparable $c/c^* \approx 2$. For direct comparison, we set a reference time $t_R$ for the time $t$ when the filament reaches $D=0.1$~mm ($\pm 10$~\textmu m). At $t\approx t_R$, all samples show a high-aspect ratio cylindrical filament characteristic of the EC regime where stretched polymer chains exert sufficient elastic stress to resist capillary pressure and retard fluid breakup relative to the Newtonian solvent. By comparing monodisperse $\lambda$ and T4 DNAs, it is evident that the longer and more flexible T4 shows the slowest filament thinning. Similarly to dilute flexible neutral polymers in low viscosity solvents, the T4-DNA shows beads on a string near the breakup time \cite{bhat2010formation,Campo2010}. Qualitatively, the filament thinning process for polydisperse ct-DNA appears to have a longer breakup time than $\lambda$ although it has a shorter peak size (i.e., $20 \times 10 ^3$ and $48.5\times 10^3$ for ct-DNA and $\lambda$, respectively). 

In Fig.~\ref{fig:2} we compare the filament diameter, $D$, and the extension rate at the neck of the filament $\dot\varepsilon = -[2/D(t)] \text{d}D(t)/\text{d}(t)$ over time ($t-t_R$) for $\lambda$, T4 and ct-DNA \cite{gier2012visualization,McKinley2005}. Generally, increasing DNA concentration leads to slower filament thinning and, for sufficient DNA concentrations, $D$ follows a decay that can be reasonably approximated as exponential for a region at $t-t_R \gtrsim 0$. The exponential region at $t\approx t_R$ marks the EC regime where a high aspect ratio cylindrical filament is developed (see Fig.~\ref{fig:Images}(d-f) and insets in \ref{fig:2}). An exponential decay of $D$ implies a constant extension rate in the EC regime as $\dot\varepsilon_{EC}=2/\tau_{EC}$. Consequently, at $t\approx t_R$ most fluids show a plateau-like region of the extension rate that differs from the monotonic increase of $\dot\varepsilon(t)$ observed for Newtonian fluids (e.g., solvent, Fig.~\ref{fig:2}). Interestingly, T4-DNAs at $c\lesssim c^*$ show a peak in the extension rate $\dot\varepsilon_p$ prior to a plateau-like region and $\dot\varepsilon_p> \dot\varepsilon_{EC}$ (Fig.~\ref{fig:2} (d)). This $\dot\varepsilon$ overshoot is typical for dilute flexible polymers in low viscous solvents and it tends to disappear as the solvent viscosity and/or the concentration of the polymer increases \cite{Rajesh2022}. Accordingly, with increasing T4 concentration, the $\dot\varepsilon$ overshoot is replaced by a smooth approach to the $\dot\varepsilon$ plateau. 

\begin{figure}[t!]
    \centering
    \includegraphics[width=8cm]{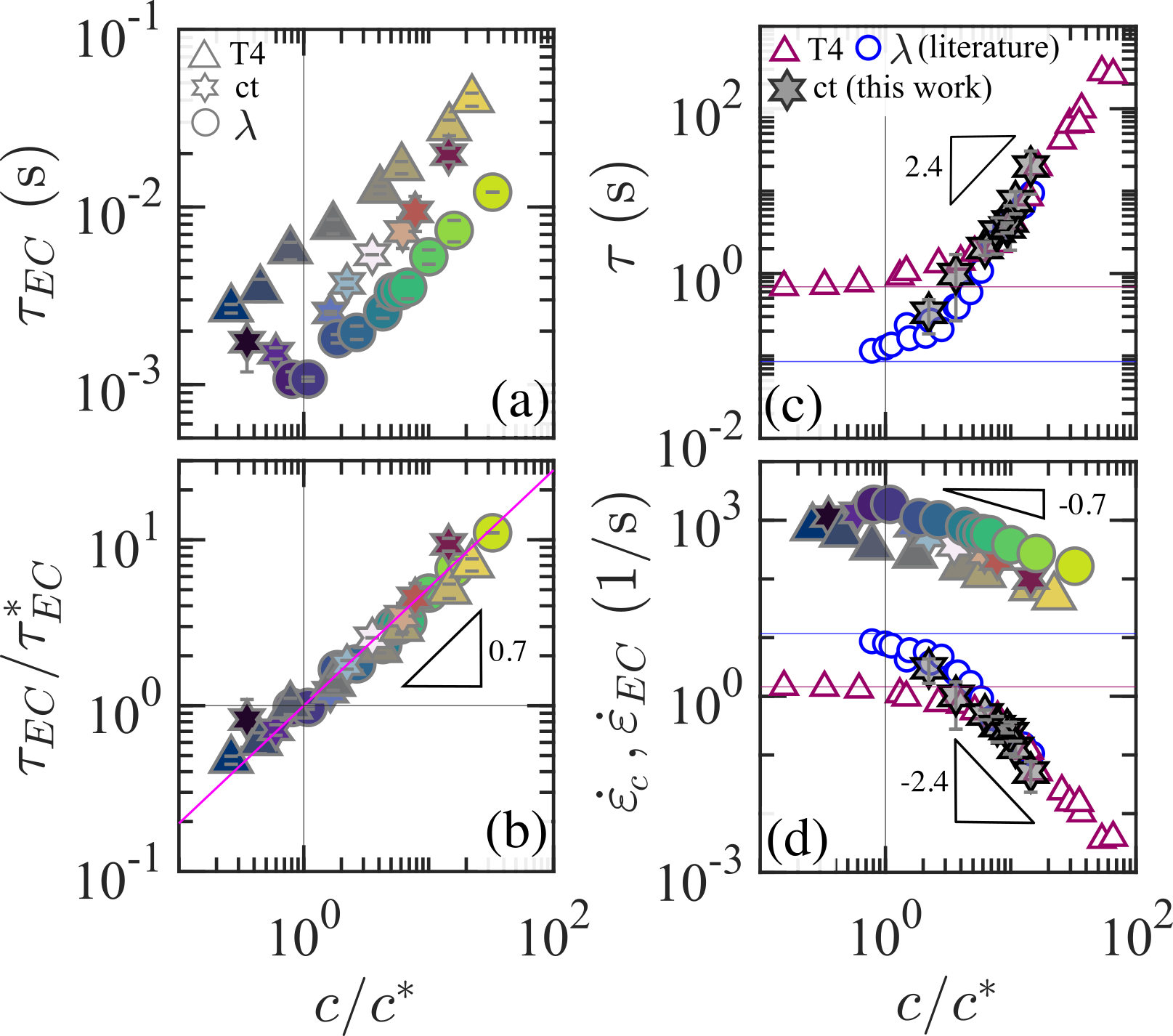}
    \caption{(a) $\tau_{EC}$ as a function of normalized concentration for $\lambda $, T4 and ct-DNA. (b) Normalized $\tau_{EC}$ with $\tau^*_{EC}$ being the value of  $\tau_{EC}$ at $c/c^*=1$. (c) Longest polymer relaxation time from rheometric measurements from this work for ct-DNA, and from literature for $\lambda$~\cite{hsiao2017direct,pan2014universal,zhou2018dynamically} and T4~\cite{liu2009concentration}. (d) Comparison of $\dot\varepsilon_c=1/\tau$ with $\dot\varepsilon_{EC}$. Horizontal lines in (c) and (d) indicate for $\lambda$ and T4 the value of $\tau$ and $\dot\varepsilon_c$ for $c\rightarrow 0$.
    Color code indicates the concentration (in \textmu g/mL) given in Fig.~\ref{fig:2}.}
    \label{fig:3}
\end{figure}

The value of $\tau_{EC}$ recovered from fitting eqn.~\ref{eqn.1} for a time span starting at $t-t_R = 0$  (see dashed lines in Fig.~\ref{fig:2}) as a function of $c/c^*$, allows direct comparison of the role of $\lambda$, T4, and ct-DNA in the filament thinning dynamics (Fig.~\ref{fig:3}~(a)). The DNAs tested show a similar power law increase of $\tau_{EC}$ with concentration. The T4 shows $\tau_{EC}$ significantly greater than that of the $\lambda$-DNA implying that the longer T4-DNA exerts greater elastic resistance to fluid breakup compared to the shorter $\lambda$-DNA.  Although ct-DNA has a peak size distribution smaller than that of $\lambda$-DNA (i.e., $20\times 10^3$ vs. $48.5\times 10^3$ bp, Fig.~\ref{fig:Images}(a-c)), ct-DNA shows a greater $\tau_{EC}$ compared to monodisperse $\lambda$-DNA, suggesting that the high-bp tail of the ct-DNA distribution contributes to $\tau_{EC}$. By normalizing $\tau_{EC}$ with $\tau^*_{EC}$, the value of  $\tau_{EC}$ at $c/c^*=1$, the curves collapse onto a master-curve (Fig.~\ref{fig:3}(b)). This indicates that $\tau_{EC}$ can be described as $\tau_{EC} =\alpha c^{\beta}$ with $\beta=0.7$ independent of DNA size and polydispersity.  Comparing $\tau_{EC}$ for $\lambda$ with the longer T4, we learn that the proportionality factor $\alpha$ increases with DNA size, thus $\tau_{EC}(c)$ for T4-DNA shifts upward compared to the smaller $\lambda$-DNA (Fig.~\ref{fig:3}(a)). 

We evaluate the polymer dynamics during the EC thinning regime by comparing $\tau_{EC}$ with the longest polymer relaxation time $\tau$. This comparison of time scales is also insightful given the customary interpretation $\tau_{EC}/3\equiv \tau$ based on the Oldroyd-B model. Using single-molecule analysis and shear rheometry, the $\tau$ dependence of $\lambda$ and T4 concentration has been reported~\cite{hsiao2017direct,pan2014universal,zhou2018dynamically,liu2009concentration}. 
For $\lambda$ and T4-DNA, we plot in Fig.~\ref{fig:3}(c) the literature values of $\tau$ considering our aqueous solvent with viscosity $\eta_s =0.9$~mPa~s. For ct-DNA, we experimentally determine $\tau$ from shear rheometry as the time scale for the onset of shear thinning (Supplemental Material \cite{ESIref}). Given the polydispersity of ct-DNA, $\tau$ from rheometry should reflect the longest relaxation time of the larger polymers in the bp-distribution.   
Comparing $\tau_{EC}$ (Fig.\ref{fig:3}(a)) and $\tau$ (Fig.~\ref{fig:3}(c)), it is evident that they are on completely different time scales and $\tau_{EC}\ll \tau$. This mismatch between time scales, which becomes even more pronounced when comparing $\tau_{EC}/3$ with $\tau$, has recently been identified as a common feature of semi-flexible polymers, as they do not conform to the assumptions of the Oldroyd-B model \cite{calabrese2024polymers}. Our experiments confirm these previous findings, but additionally indicate that $\tau_{EC}$ does not reflect the scaling of $\tau$ with concentration. In fact, $\tau_{EC}$ appears to follow the same scaling with concentration regardless of the transition between concentration regimes at $\approx c^*$ (dilute to semidilute) and $\approx 3c^*$ (semidilute to entangled). At $c/c^* \gtrsim 10$, all DNAs follow a similar $\tau \sim c^{2.4}$ scaling while $\tau_{EC}\sim c^{0.7} $, resulting in a deviation of time scales increasing with concentration as $\tau/\tau_{EC} \sim  c^{1.7}$.
The fact that $\tau_{EC}\ll \tau$ also implies that $\dot\varepsilon_{EC} $ is greater than the critical extension rate required for polymers to stretch $\dot \varepsilon_c \approx 1/\tau$.  The comparison of $\dot\varepsilon_{EC}$  with $\dot \varepsilon_c$ (Fig.~\ref{fig:3}(d)) shows that depending on polymer concentration, $\dot\varepsilon_{EC}$ is $10^2$ to $10^4 \times$ greater than $\dot \varepsilon_c$, suggesting that the EC response is ruled by highly stretched polymers \cite{sattler2007molecular, calabrese2024polymers,juarez2011extensional}. Thus, $\tau_{EC}$ can be considered as a time scale informative of DNAs far from equilibrium.

Measurements of $\lambda$-DNA and polyelectrolytes at high deformation rates (i.e., when chains are highly stretched) suggested that stretched polymers under strong flows are unable to perceive the confinements imposed by the surrounding stretched chains even at concentrations exceeding $c^*$ \cite{dakhil2019infinite,dakhil2021buffered}. This led to the concept that highly stretched semi-flexible polymers can experience a dilute-like environment even at $c>c^*$ \cite{hsiao2017direct,dakhil2019infinite,dakhil2021buffered}. The scaling $\tau_{EC} \sim c^{\beta}$ persisting over three different concentration regimes also supports the picture that highly stretched DNAs are insensitive to interpolymer interactions. At least for the model monodisperse DNAs, the occurrence of interpolymer interactions would have been expected as a deviation of the power exponent $\beta$ at $c\approx c^*$ and $c\approx 3c^*$ (akin to $\tau$ and zero-shear rate viscosity).

A natural  question is why does the polydisperse ct-DNA follow the same $\tau_{EC}\sim c^\beta$ scaling as the monodisperse DNAs? To answer this question, we consider polydisperse ct-DNA as a distribution of $n$ monodisperse DNA fractions that actively contribute to the elastic stress during the EC regime. Each DNA fraction has a concentration $c_i$, a proportionality factor $\alpha_i$ and a characteristic time scale $\tau_{ECi}=\alpha_i c_i^\beta$. Neglecting interpolymer interactions, the preserved $\tau_{EC}\sim c^\beta$ scaling for the polydisperse DNA can be hypothesized if $\tau_{EC}$ is a summation of each $\tau_{ECi}$ contribution for a constant $n$ DNA fractions as
\begin{equation}
    \tau_{EC}(c)=\sum_{i=1}^{i=n} \tau_{ECi}=\sum_{i=1}^{i=n} \alpha_i c_i^\beta.
    \label{eqn.Summation}
\end{equation}
Since also for ct-DNA $\dot\varepsilon_{EC} \gg \dot\varepsilon_c$, at least the high-bp fraction should be stretched and contribute to the EC response. 
\begin{figure}[t]
    \centering
    \includegraphics[width=8.5cm]{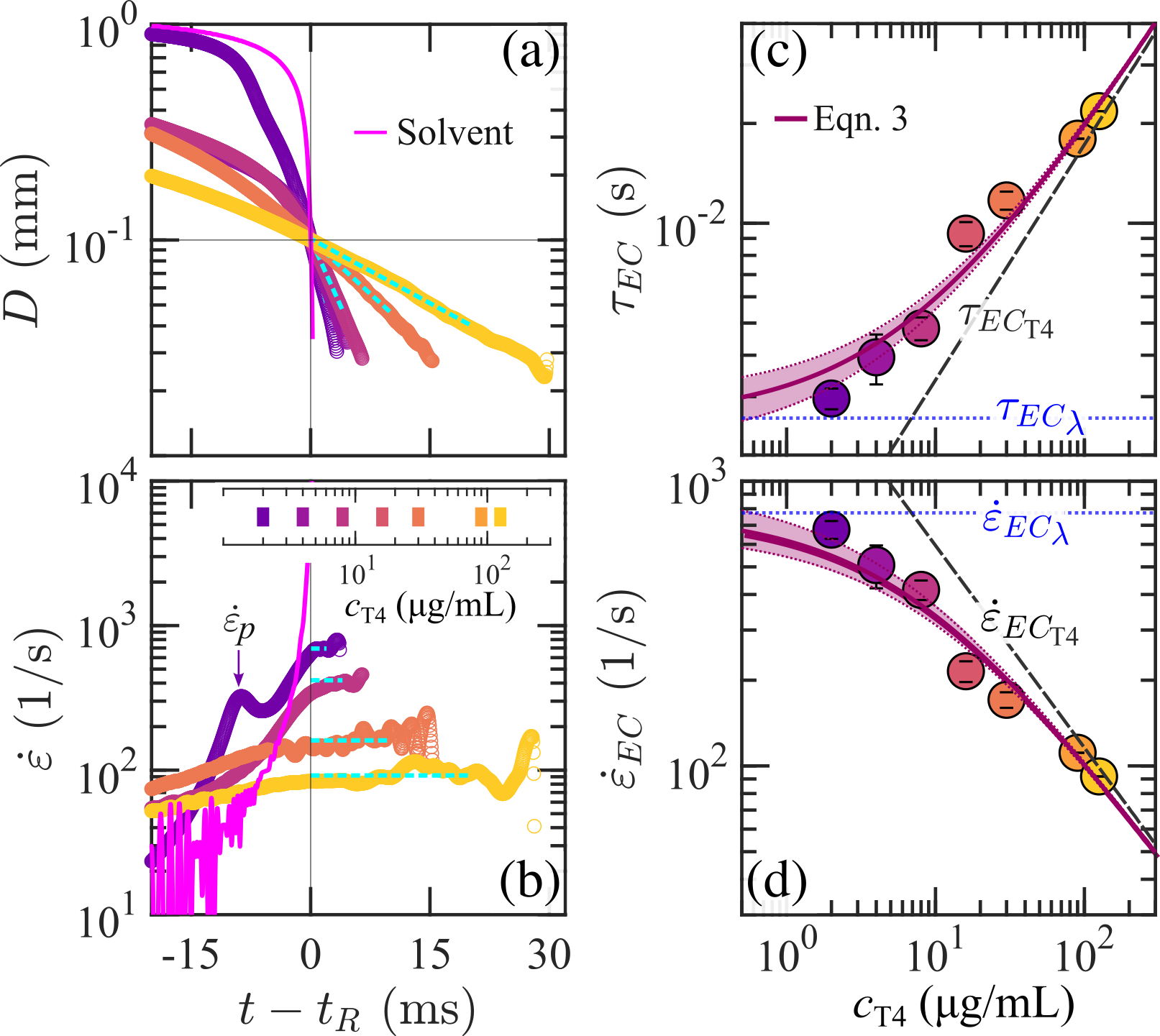}
    \caption{(a) Filament diameter $D$ and (b) extension rate $\dot\varepsilon$ as a function of time $t-t_R$ for bidisperse solutions composed of a background $\lambda$-DNA concentration ($c_{\text {\small $\lambda$}}=202$~\textmu g/mL) and varying T4 concentration $c_{\text{\small T4}}$. The dashed lines in (a) and (b) indicate the fit to eqn. \ref{eqn.1} the corresponding $\dot\varepsilon_{EC}$, respectively.(c) $\tau_{EC}$ and (d) $\dot\varepsilon_{EC}$ as a function $c_{\text{\small T4}}$, respectively. The shaded region indicate the model uncertainty based on the error of $\tau_{EC _{\text{\small $\lambda$}}}$. In (c) and (d), the dashed black line indicates the trend of the monodisperse $\tau_{EC _{\text{\small T4}}}$ (or $\dot\varepsilon_{EC_{\text{\small T4}}}$) and the dotted blue line indicates $\tau_{EC_{\text{\small $\lambda$}}}$ (or $\dot\varepsilon_{EC_{\text{\small $\lambda$}}}$) of the background $\lambda$-DNA solution.}
    \label{fig:BiDisperse}
\end{figure}
To test this hypothesis, we formulate bidisperse DNA solutions composed of mixtures of $\lambda$ and T4 with established relationship between $\tau_{EC}$ and their respective concentrations (Fig.~\ref{fig:3}(a, b)). We blend a fixed background concentration of $\lambda$-DNA ($c_{\lambda}=202$~\textmu g/mL) with a measurable $\tau_{EC}$ on its own, with T4-DNA at different concentrations. Qualitatively, the thinning dynamics of the bidisperse solutions (Fig.~\ref{fig:BiDisperse} (a ,b)) are analogous to the mono- and polydisperse cases described in Fig.~\ref{fig:2}. Interestingly, the relatively small addition of T4 in the mixture leads to a peak in the extension rate $\dot\varepsilon_p$ (Fig.~\ref{fig:BiDisperse}(b)), akin to that observed for pure T4 shown in Fig.~\ref{fig:2}(d). However, for the bidisperse solution $\dot\varepsilon_p<\dot\varepsilon_{EC}$, an opposite feature to that observed for the pure T4 solution (Fig.~\ref{fig:2}(d)). This suggests that in the bidisperse solution, as the capillary thins and $\dot\varepsilon$ increases, the T4 population stretches, giving rise to the characteristic peak in the extension rate $\dot\varepsilon_p$, prior to the stretching of the smaller $\lambda$. 
From the hypothesis that $\tau_{EC}$ can be described by a summation of $\tau_{EC i}$  contributions of each DNA fraction we recast eqn.~\ref{eqn.Summation} to describe the bidisperse DNA system as 
\begin{equation}
      \tau_{EC}(c_{\text{\small T4}})=\tau_{EC_{\text{\small $\lambda$}}}
+\tau_{EC_{\text{\small T4}}}= \alpha_{\text{\small $\lambda$}} c_{\text{\small $\lambda$}}^\beta+ \alpha_{\text{\small T4}} c_{\text{\small T4}}^\beta,
      \label{eqn.bidisperse}
 \end{equation}
where the subscripts refer to the terms describing the $\lambda$ and T4 fractions. Since the functions describing $\tau_{EC_{\text{\small $\lambda$}}}(c_{\text{\small $\lambda$}})$ and $\tau_{EC_{\text{\small T4}}}(c_{\text{\small T4}})$ for pure DNAs are known (presented in Fig.~\ref{fig:3}(a)), $\tau_{EC}$ predicted by eqn.~\ref{eqn.bidisperse} can be directly compared with our experimental results. The experiments and the summation model show that with increasing $c_{\text{\small T4}}$ in a constant background solution of $\lambda$-DNA, $\tau_{EC}$ rises above the background value of $\tau_{EC_{\text{\small $\lambda$}}}$ and at high concentrations of $c_{\text{\small T4}}$, $\tau_{EC}\rightarrow $ $ \tau_{EC_{\text{\small T4}}}$.

Our data and model indicate that for semi-flexible DNA, $\tau_{EC}$ is not determined by the longest $\tau_{ECi}$ but corresponds to the summed contributions of $\tau_{ECi}$ from each polymer fraction. The fact that $\dot\varepsilon_{EC}$ is at least $10^2 \times$ greater than the respective $\dot \varepsilon_c$ of each monodisperse DNA solution shown in Fig.~\ref{fig:3}(d) supports the idea that both polymer fractions contribute to the EC response by being highly stretched. This behavior contrasts sharply with that of flexible polymers, where the stretching of a high-$Mw$ polymer fraction can result in a $\dot\varepsilon_{EC}$ insufficient to stretch the low-$Mw$ polymers \cite{calabrese2024effects}. 
This difference between flexible and semi-flexible polymers is a clear indication of the greater impact exerted by flexible polymers in the flow. It is important to note that given the relatively high background concentration of $\lambda$-DNA ($5c^*$), all bidisperse solutions are expected to exhibit strong interactions at equilibrium. Regardless, our simple model neglecting interpolymer interactions, captures the $\tau_{EC}$ (and $\dot\varepsilon_{EC}$) trend. This supports the conjecture that DNAs (and presumably semi-flexible polymers in general) under strong flows interact less than when at equilibrium. 

Given the continuous integration of \textit{rheological} techniques based on capillary-driven thinning in various fields (e.g., refs.~\cite{bugarin2024rheological, marconati2020role, ruhs2021complex}), and the crucial role that capillary-driven thinning and breakup play in both industrial and biological contexts, there is increasing research dedicated to deepening our understanding of this process (e.g., refs.~\cite{McKinley2005, Campo2010, dinic2020flexibility, Gaillard2023, gaillard2024does, zinelis2024fluid, aisling2024importance}). We have systematically investigated the behavior of model mono-, bi-, and polydisperse semi-flexible polymers in capillary-driven extensional flows. Our findings reveal that the elastocapillary time constant $\tau_{EC}$ does not directly correlate with the longest polymer relaxation time, as traditionally assumed. We propose and validate a predictive model that accounts for the time scale contribution of each polymer in the distribution. The validity of this model suggests that for semi-flexible polymers,  strong extensional flows lead to decreased interpolymer interactions. We believe that these results will motivate direct evaluation of polymer dynamics in capillary-driven flow and support the development of mathematical models to predict the elastocapillary time constant $\tau_{EC}$ from known polymer parameters.

The authors gratefully acknowledge the support of Okinawa Institute of Science and Technology Graduate University (OIST) with subsidy funding from the Cabinet Office, Government of Japan. We are grateful for the help and support provided by the Sequencing Section of Core Facilities at OIST. V.C., S.J.H., A.Q.S. also acknowledge the financial support from the Japanese Society for the Promotion of Science (JSPS, Grant Nos. 24K07332, 24K17736, and 24K00810). 


%

\end{document}